# Data Leaves
## Scenario-oriented Metadata for Data Federative Innovation

Yukio Ohsawa[1], Kaira Sekiguchi[1], Tomohide Maekawa[2],
Hiroki Yamaguchi[2], Son Yeon Hyuk[1], Sae Kondo[3]

[1] The University of Tokyo, [2] Trust Architecture, Inc., [3] Mie University

Abstract: A method for representing the digest information of each dataset is proposed, oriented to the aid of innovative thoughts and the communication of data users who attempt to create valuable products, services, and business models using or combining datasets. Compared with methods for connecting datasets via shared attributes (i.e., variables), this method connects datasets via events, situations, or actions in a scenario that is supposed to be active in the real world. This method reflects the consideration of the fitness of each metadata to the feature concept, which is an abstract of the information or knowledge expected to be acquired from data; thus, the users of the data acquire practical knowledge that fits the requirements of real businesses and real life, as well as grounds for realistic application of AI technologies to data.

## 1. Introduction: feature concepts as representation of scenarios

As discussed in the context of chance discovery [1], the consideration of cross-points of multiple scenarios provides clues for decision-making in a dynamic environment. A scenario here refers to a sequence of events, situations, and/or actions connected in a certain context. Various scenarios appear in different event domains, and some scenarios appear across multiple domains. For example, in marketing, as in Fig.1(a), a customer of a supermarket may come in to buy beer and become interested in a new beer. However, after buying the beer, by searching the brand name of this beer using on the Web, he found that the price of the beer in this store is 30% more expensive than in other shops. Consequently, this customer refunds the beer and buys something else. To explain this transition, we need data from multiple domains, that is, the data on purchases in the supermarket and the information he touched, because this customer's interest changed by touching information out of the store.  On the other hand, in the example of document indexing, as shown in Fig.1(b), basic knowledge or established concepts are positioned based on the contextual flow in the document. New ideas may be proposed based on connecting these bases to appear as infrequent keywords that connect the bases. To explain this context flow in a document, we need to visualize the document in the metaphor of architecture, that is, building construction, as in the original publication of KeyGraph [2], and extract keywords regardless of their low frequency.

   Such an abstract sketch of the target system, which one considers (to explain the present state, predict the behaviors, etc.) using certain dataset(s), was defined as a feature concept (FC) by the author [3]. That is, an FC is the abstraction of the concept or knowledge expected to be acquired from data, capturing the features of the purpose of the data user(s), and positioning the dataset(s) that can be used to satisfy the purpose. An FC can be illustrated as an image, as shown in Fig. 1(a) or Fig. 1(b), even if the same algorithm is applied for visualization (KeyGraph; {a} corresponds to applications to marketing as in [1] and (b) to document indexing as in [2]). The components of the FC are listed below. In summary, these are similar to real-world scenarios, which may be included in the data because FC is a tool for representing considerable scenarios linking them to datasets.

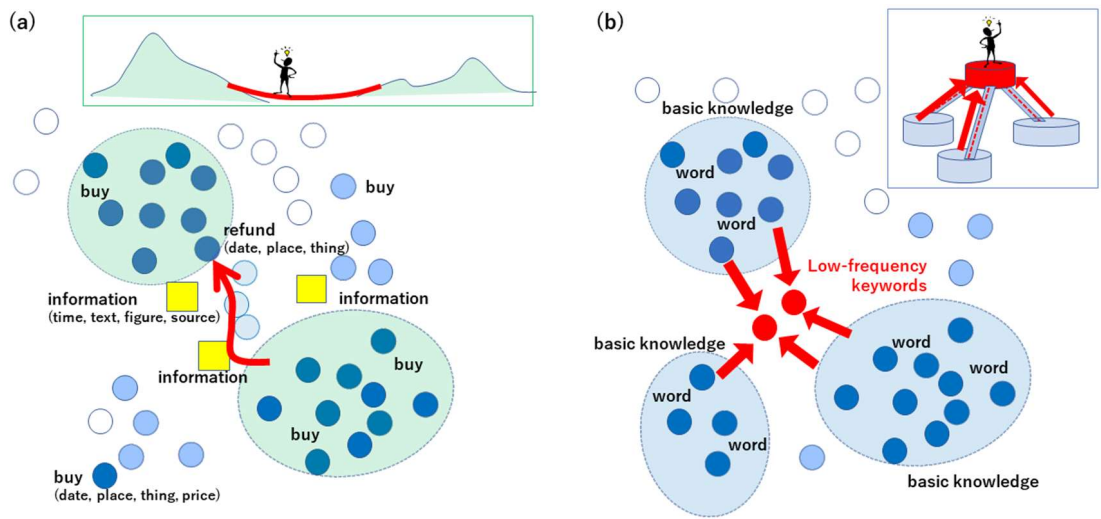

Fig.1 Feature concepts (a) islands and bridges in the market (b) architecture of words

**[The minimum set of components of a FC]**

FC component A: The ID number of the FC

FC component B: The title and abstract of the system represented by the FC

FC component C: Nodes representing events, situations, entities, or actions

FC component D: Directed (i.e., arrows) or nondirected edges representing the causality, order, continuity, or relevance of the elements represented by nodes.

FC component E: Space among elements, that is, nodes or edges, where the distance represents the relevance of the elements.

FC component F: Frames surrounding all components above, each corresponding to a context. The entire FC is regarded as a top-level frame that corresponds to the purpose of the data user.

## 2. Data Leaves: Scenario-oriented metadata
### 2.1 Metadata should fit the target system

Metadata is the data about a certain dataset. By this definition, it can be the dataset itself; however, it is normally used to refer to some digest information to explain the summary of a dataset.

Metadata is used not only for data retrieval but also for the interoperability of data from multiple domains, that is why standardization has been regarded as an important issue in designing metadata [4,5]. Typically, metadata include the title, abstract, variables that are attributes of values in the data, and the meaning of the variables. In the prevalent metadata used in data catalogues, the variables are regarded as information for which the scientific meaning can be shared by the provider and users (real or latent) of the data. In this sense, it is normal to regard variables and information relevant to them as an essential part of metadata, as in [6,7]. However, it is difficult to include the context of the dataset in the shear variable set. For example, the variables of the point of sales (POS) dataset may include {customer ID, date, time} as a subset, a blood test dataset may have {patient's name, test date}, and the government has a pairwise set {name, my number}. We may be able to relate the customer ID and the patient's name via governmental data if we have a correspondence [customer ID, name]. However, as far as we do not have such a piece of information implying "customer" is one of the people in some countries, there is no way to relate the data of POS and blood tests, although this combination can be useful, as in the example in Fig.1. Similarly, {item} in the POS data and {tested item} in the blood test may be linked directly if the computer highlights the "item" as the common word, which is meaningless as far as we miss the bridging information between them such as "item" – "food to eat" – "nutrition" – "blood condition" – "tested item." Thus, we need information about events, actions (e.g., eating), situations (e.g., blood conditions), or entities (e.g., food, people) in the metadata to show meaningful connectivity and interoperability among datasets. Some existing metadata explicitly include such elements [8-11], while others allow inclusion in the abstract, as in the data jackets (DJs) below [12]. Furthermore, others proposed the representation of the logical relation between variables and entities and intended to execute reasoning on knowledge [9, 10], which is not designed to reflect the purpose of wrapping a feature concept that may be a loose combination of various pieces of logical knowledge, some parts of which can be clarified only after the data are analyzed or visualized for interpretation by humans (s).

## 2.2 Data Jackets and Innovators Marketplace on Data Jackets

A DJ is metadata for each dataset, reflecting subjective or potential interest in the data market. A DJ includes the title, abstract, attributes (variables in the data), and expectations about the utility of the corresponding dataset that may be included in the abstract. The expectation can be written subjectively, showing the desires of data users. By visualizing the relevance of DJs, participants in the data market think and talk about why and how to combine the corresponding datasets. Even if the owners of data may hesitate to open their data to the public, they can present only the DJs to the public society because they do not include the content of the data. The Innovators Marketplace on Data Jackets (IMDJ) is a platform for data-federative innovations; that is, an environment in which participants communicate to create ideas for combining/using/reusing data or future collaborators. In the snapshot in Fig. 2, referring to the map visualizing the connectivity of datasets using DJs, the participants communicate requirements and solutions to the requirements. The ideas regarding solutions by Dr. Y, that is, how datasets can be used to satisfy the requirement by Dr. X, can be proposed on DJs, the small pieces of information summarizing the datasets and their utility, connected, and visualized in the map. Data user Dr. X or analyst Dr. Y may pay the provider for the data (Dr. Z).

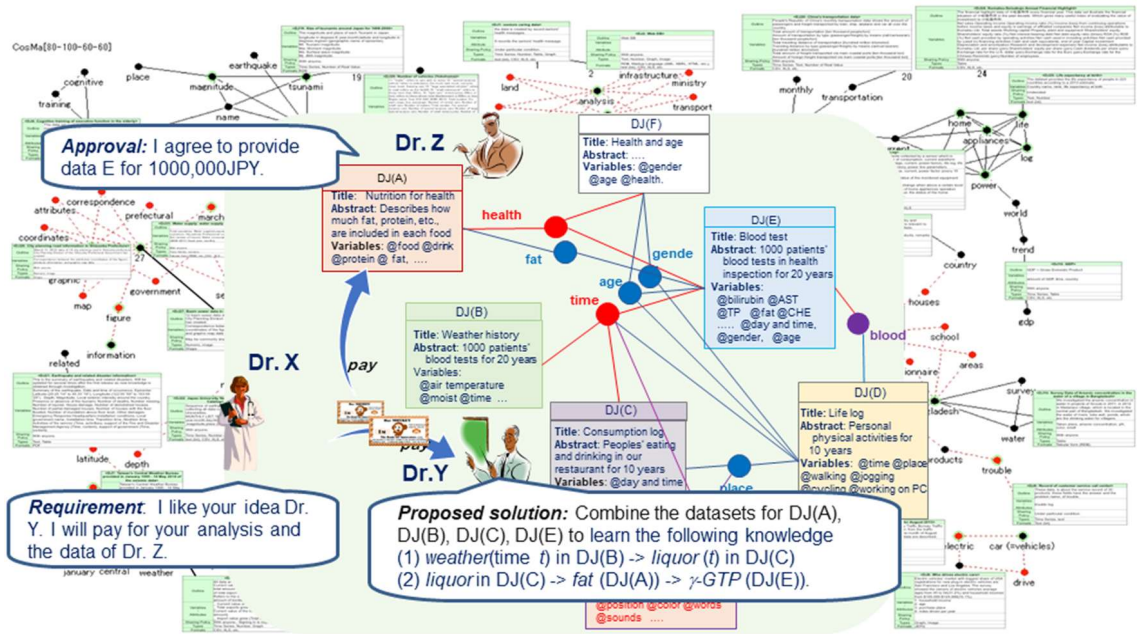

Fig.2 A snapshot of communication in IMDJ, with a map of connected Data Jackets (the small cards in the background, with the enlarged abstraction in the front).

## 2.3 Data Leaves and their structure

A data leaf is a metadata designed to wrap a feature concept. A data leaf (DL) includes the following information so that a set of DLs corresponding to a set of datasets fits the FC.

  DL_info A: The ID number of the data leaf

  DL_info B: The title and the abstract of the data represented by the DL

  DL_info C: The events, situations, entities, or actions in the system that the dataset is about.

  DL_info D: Directed (i.e., arrows) or nondirected edges representing the causality, order, continuity, or relevance of the elements in DL_info C

  DL_info E: The space among elements, that is, nodes or edges, where the distance represents the relevance of elements according to the data corresponding to the DL

An FC comes to be wrapped by DLs in a co-evolutionary procedure in which the transformation of the FC and the development of DLs occur co-operatively, as shown in Fig.3. The upper half of the figure is equivalent to the FC in Fig.1(a), which is wrapped by the two DLs that evolve during the creation of the FC. The FC can be revised to reflect the DLs, which can also be revised from the initial structure to (1) fit the FC and (2) reflect the results of the analysis of the corresponding datasets using AI technologies. The overall process of the FC-DL co-evolution proceeds as follows.

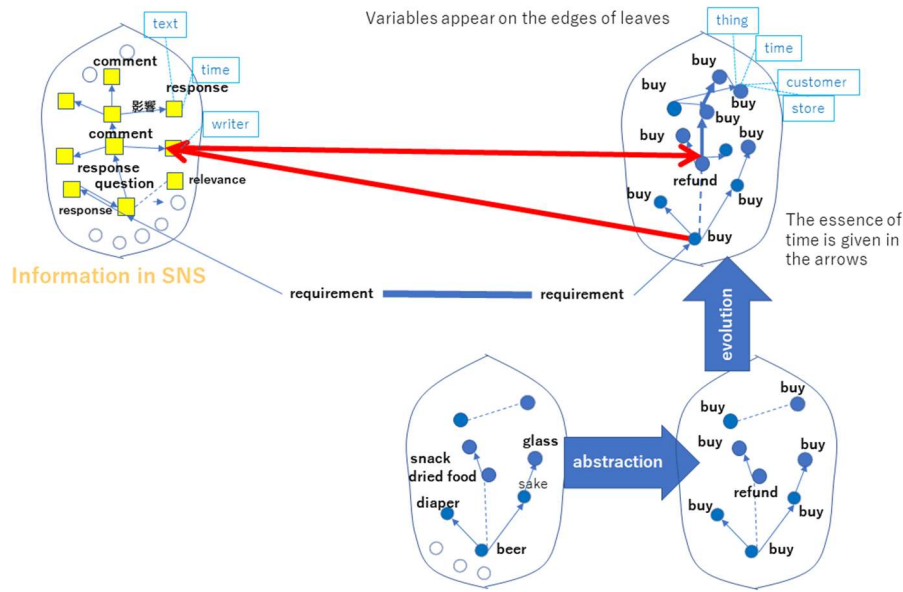

Fig.3 Two DLs fitting the feature concept "islands and a bridge in the market." The DLs are developed from the initial structure for two purposes: (1) fitting the FC and (2) reflecting the results of the analysis of the corresponding datasets

**The procedure of FC-DL co-evolution**

**0. Prepare DLs**

Register used DLs. In each DL, events, situations, and actions are represented by nodes.
The temporal order or causality and the relevance among nodes are represented by arrows and lines or distances among nodes.

**1. Creation of FC ver. 0**

- Events, situations, and/or actions in the target system are represented by nodes.
- The temporal order, causality, and relevance among nodes are represented by arrows and lines or distances among nodes, respectively.

**2. Divide FC into islands**

- Group by major categories/categories/themes/etc. to make islands
- and connect islands with arrows if there is a causal relationship, etc.

**3. Re-structure the FC ver.0 into ver.1:**

- DLs were used to wrap the FC and fit the nodes to the corresponding nodes in the FC.
- Variables in the data are indicated by edge of the DL.
- Events and elements that span multiple DLs are placed between leaves to form bridges.

- Causality and temporal orders are indicated by arrows.
- The positions of the nodes and edges can be changed as long as the structure, that is, the order, causality, or relevance among the elements of the FC, is sustained.

4. **additional DL registration:** Create and register additional DLs missed in the above 3.

## 2.4 A way to translate variables in a DJ to elements in a DL

In some cases, DLs are desired to be created by translating them from the existing DJs. As shown in Fig. 4, the variables in a DJ are replaced with a representation of causalities or scenarios, including elements of the real world such as events, situations, entities, or actions. By connecting these elements by arrows representing causalities or orders of occurrence, as shown in the upper figure, the author of DL becomes aware of latent elements and adds them to the represented scenarios. To simplify the task of this translation, the choice is to categorize the variables into levels of functionality. In the example shown in Fig.4, "cold"/"hot" represents a tentative state to be expressed as a high/low value of the variable "temperature." On the other hand, the variable "temperature" can be also regarded as a function having "time" and "place, " etc.. Such functional variables can be used as elements of DL by interpreting it as a verb (action), adjective (situation), or an event as a change in the situation on the side of the user.

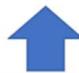

Fig.4 The creation of a DL, transformed from a corresponding DJ

## 3. Experimental IMDL: IMDJ using Data Leaves

### 3.1 Settings

We conducted experiments on IMDJs using data leaves (IMDLs) in two settings. In setting (a), we used the original DJs, as shown in the lower half of Fig.4. In setting (b), we used and the created DLs in the upper half of Fig.4 as DJs.

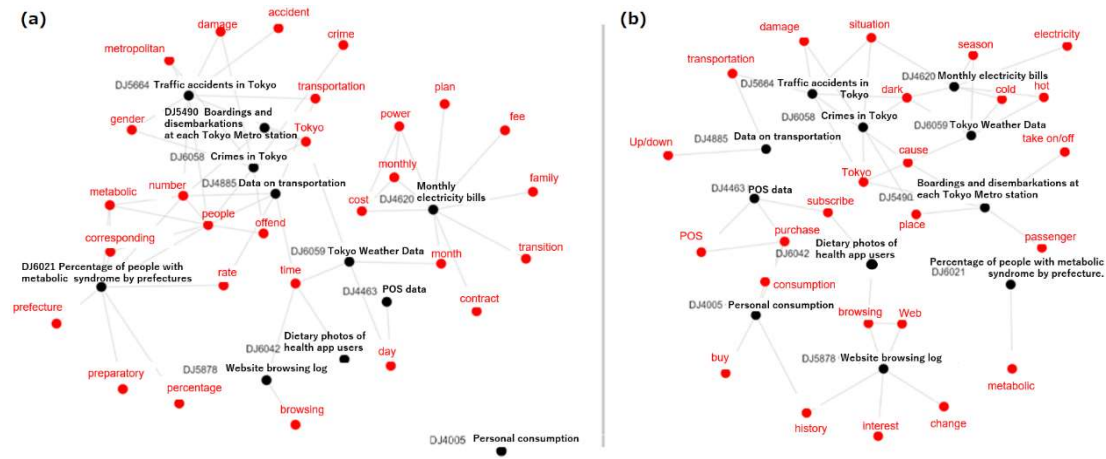

Fig.5 Two maps used in IMDL on KeyGraph using original (a) DJs (b) DLs

We had ten participants here, five of whom played (a) for 15 min, followed by (b) for 15 min. The other five participants played (b) for 15 minutes, followed by (a) for 15 minutes. For all 30 min, requirements and solutions were put on the same maps of KeyGraph in Fig.5, that is, maps (a) and (b), respectively, received stickers from both groups. Finally, we counted the number of requirements and solutions.

### 3.2 Results

As a result, DLs are found to be significantly preferred to DLs as a tool for innovative thoughts and communication, according to the result of "good to use" as in Table 1 although the difference in the number of proposed requirements and solutions was not significant. According to common user comments, the reasons for this preference are threefold. First, the DLs on which the participants could create solutions were highlighted in the map by connecting to a comparably large number of links. For example, the DL corresponding to "DJ6059 Tokyo Weather Data" in (b) is connected to five, a larger number than four in (a), words connected to "cold" "hot" "dark", which show the situational information of habitants in each region at each time. In contrast, "Tokyo," "time," "day," and "month" connected to DJ6059 in (a) are neither related to events or situations. The other reason was that it was easy to think of scenarios using the "red" words in the map, such as "hot" and "cold" describing situations and "consumption" and "purchase" to describe actions. Third, it was easy to create data-use scenarios that inherited what others had written.

Table 1: Quantitative comparison of IMDJ (using DJs) and IMDL (using DLs)

| The map is on / Number of | (a) IMDJ | (b) IMDL |
|---|---|---|
| Requests (yellow stickers) | 16 | 20 |
| Solutions (blue sticker) | 20 | 19 |
| Better for creating solutions | 2 | 8 |

## 4. Discussions

The above results of the IMDJ imply the possibility that showing events relevant to a scenario underlying data plays a role in discovery and knowledge acquisition. This effect originates from the visualized words representing events, situations, entities, or actions that connect datasets according to the experiment according to the reasons above, mentioned by the participants in the experiment. Furthermore, the positions of the requirements (yellow stickers) and the solutions (blue) tend not to appear close to the non-functional variables such as 月 (month), 時(time), 性別 (sex), 数(number). Thus, we converge on the tendency that components of DLs reinforce human thoughts in data federative innovation. The reasons for the users' preferences discussed here imply that words corresponding to events, situations, and actions associate human thought with parts of the FC, which can be regarded as an abstraction of the dynamics of the target system or the considered part of the real world that can be regarded as an abstraction of the dynamics of the target system or the considered part of the real world which bundles scenarios, that is, sequences of events, situations, and actions, of to be realized by data federation. In this sense, the connectivity among datasets via the elements of these scenarios is expected to compose useful combinations of data. Thus, tools designed on DJs, developed so far [13, 14], are expected to improve better in aiding data federative innovations by replacing DJs with DLs.

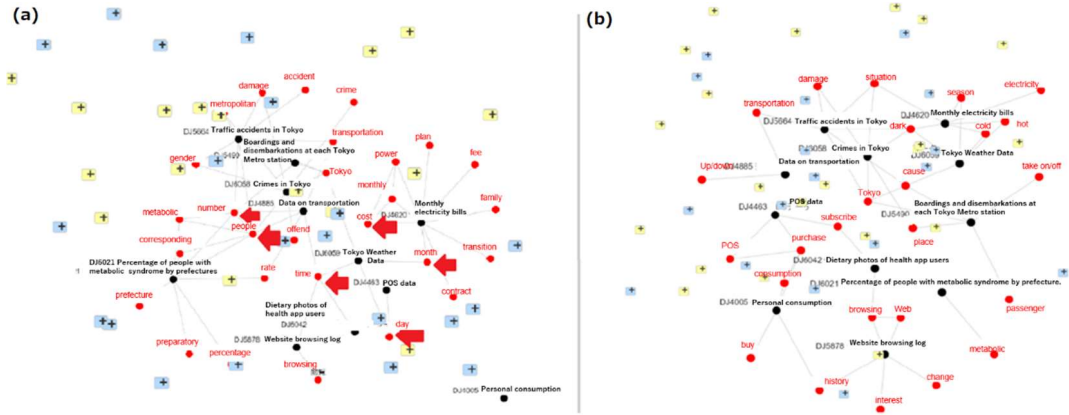

Fig.6 The distribution of requirements and solutions on the maps (a) DJs (b) DLs: Red arrows show the variables which are not functionable. Requirements and solutions are sparse close to the red arrows even in (a).

## 5. Conclusions

The data leaf is named after its structure, which includes causalities of events and its role in wrapping the FC. It should initially have a simple structure, which is a hypothetical causal network that is compared to the veins of a leaf and revised to fit the FC by learning from data that refines the fine structure of the veins. Although the central aim of using DLs is to accelerate the process of creating valuable solutions in data federative innovation, the ability of leaf veins to evolve with both the social requirements and content of data is essential. Here, the data ecosystem refers to the organic growth of the inside and the interlinks of knowledge about and from the data.

## Acknowledgment

We appreciate the members of the DFIL social collaboration course and the Ohsawa Lab. To discuss the most essential steps in inventing DLs. This study was supported by JST Grant NumberJPMJPF2013, JST Q-Leap JPMXS0118067246, JSPS Kakenhi 20K20482, and the MEXT Initiative for Life Design Innovation.